\documentclass[11pt]{article}
\usepackage{hyperref}
\pdfoutput=1 
\usepackage{titling}
\begin{document}
\title{Impinging Jet Dynamics}
\author{Xiaodong Chen and Vigor Yang \\
\\\vspace{6pt} School of Aerospace
Engineering, \\ Georgia Institute of Technology, Atlanta, GA 30332, USA}
\maketitle
Collision between two cylindrical jets is one of the generic configurations for the generation of liquid sheets, 
the dynamics and stability of which have attracted a great deal of attention due to their relevance to the spray atomization
 and combustion process. In the present work related to this fluid dynamics video, high fidelity numerical simulations 
 were performed to study the dynamics and stability of impinging jets over a broad range of operating conditions. An improved volume-of-fluid (VOF) method augmented with an innovative topology-oriented adaptive mesh refinement (TOAMR) technique was used to simulate the formation and breakup of the liquid sheet formed by two impinging jets.  The behaviors in various Reynolds and Weber number regimes are studied systemically.  The predicted liquid sheet topology, atomization, and droplet size distribution agree well with experimental measurements. 

In this fluid dynamics video, Ray-tracing data visualization technique was used to obtain realistic and detailed flow 
motions during impinging of two liquid jets. Different patterns of sheet and rim configurations were presented 
to shed light into the underlying physics, including liquid chain, closed rim, open rim, unstable rim and flapping sheet. 
In addition, stationary asymmetrical waves were observed and compared with existing theories. The generation of 
stationary capillary wave in respect to the liquid rim were explained by the classic shallow water wave theory. 
The atomization process caused by development of the impact waves were observed in detail, including fragmentation of 
liquid sheet, formation of liquid ligaments, and breakup of ligament into droplet. The locking-on feature of the
wavelength of impact wave were also found to be similar to that of perturbed free shear layers.
\end{document}